\documentclass[epsfig,pre,preprint,superscriptaddress]{revtex4}
\usepackage{graphicx}
\usepackage{dcolumn}
\usepackage{amsmath}    
\usepackage{amssymb}
\usepackage{bm} 
\usepackage{hyperref}
\usepackage{latexsym}
\usepackage{verbatim}
\usepackage{color}
\usepackage{subfigure}
\usepackage{floatflt,graphicx}
\usepackage{appendix}
\def\beq{\begin{equation}}
\def\eeq{\end{equation}}

\def\etal{{\it et al.}}

\def\Q{\mbox{\sffamily\bfseries Q}}

\def\beq{\begin{equation}}                           
\def\eeq{\end{equation}}                           
\def\bea{\begin{eqnarray}}                           
\def\eea{\end{eqnarray}}        
\def\nhat{\hat{\bf n}}

\draft
                   
\begin{document}
\textwidth = 6.7 in
\textheight = 9 in
\oddsidemargin = 0.0 in
\evensidemargin = 0.0 in
\topmargin = 0.0 in
\headheight = 0.0 in
\headsep = 0.0 in
\parskip = 0.2in
\parindent = 0.0in

\title{A Dynamic Renormalization Group Study of Active Nematics}
\author{Shradha Mishra}
\email{smishr02@syr.edu}
 \affiliation{Physics Department, Syracuse University, Syracuse NY 13244 USA}
\author{R. Aditi Simha}%
 \email{aditi@physics.iitm.ac.in}
\affiliation{Department of Physics, Indian Institute of Technology Madras, 
Chennai 600 036, India}
\author{Sriram Ramaswamy}%
 \email{sriram@physics.iisc.ernet.in}
\affiliation{Centre for Condensed Matter Theory, Department of Physics, Indian Institute of Science, 
Bangalore 560 012 India}

\date{\today}

\begin{abstract}
{We carry out a systematic construction of the coarse-grained dynamical equation of motion for the orientational order parameter for a two-dimensional active nematic, that is a nonequilibrium steady state with uniaxial, apolar orientational order. Using the dynamical renormalization group, we show that the leading nonlinearities in this equation are marginally \textit{irrelevant}. We discover a special limit of parameters in which the equation of motion for the angle field of bears a close relation to the $2d$ stochastic Burgers equation. 
We find nevertheless that, unlike for the Burgers problem, the nonlinearity is marginally irrelevant even in this special limit, as a result of of a hidden fluctuation-dissipation relation. 
$2d$ active nematics therefore have quasi-long-range order, just like their equilibrium counterparts}
\end{abstract}

\maketitle

Active nematics \cite{sradititoner,chateginellimontagne,vjmenonsr} are the
simplest example of spontaneously broken rotation-invariance in a
nonequilibrium system. Analytical studies of their statistical properties have
mainly been confined to a linearized approximation \cite{sradititoner}, whose
predictions of anomalous density fluctuations have largely been confirmed in
experiments \cite{vjmenonsr} and numerical simulations
\cite{chateginellimontagne}. Within the theory of \cite{sradititoner} the
density fluctuations were driven by the broken-symmetry modes associated with
orientational order. In this paper we ignore density fluctuations and focus on
the effect of the broken-symmetry modes on the strength of orientational order.
We ask: can a noisy two-dimensional system of active particles display
{\em long-range} nematic order?

Let us see why this question is worth asking. It is well known that {\em at thermal
equilibrium}, in two space dimensions, neither XY models nor nematic liquid
crystals can have long-range order. Instead of a true ordered phase, these
systems have a critical low-temperature state in which the fluctuation-averaged
order parameter vanishes in the thermodynamic limit at all nonzero
temperatures, but order-parameter correlations decay as a power of distance
\cite{MW,hohenberg,KT,frenkels}. The simplest generalization of the $2d$ XY
model to a nonequilibrium steady state is the Vicsek model \cite{vicsek} of
flocks in two dimensions, in which the local velocity of the flock is the XY order-parameter
field. Toner and Tu \cite{tonertu} showed that the resulting advection of the 
order-parameter field by its own fluctuations \cite{tonertu} {\em stabilizes} 
long-range order even in two dimensions. Technically, the mechanism amounted to a singular
renormalization of the XY stiffness by nonlinearities of a type not permitted
in the equilibrium XY model. The ordered state of a Vicsek flock can be thought
of as a collection of arrows all pointing on average in the same direction;
this is known as {\em polar} order. One can imagine a different ordered state,
in which the {\em axes} of the arrows are on average parallel to an arbitrarily
chosen spatial direction, call it $\nhat$, but the arrows point indifferently
along $+\nhat$ and $-\nhat$ or,  equivalently, one could simply lop the heads
off the arrows.  The resulting state is {\em apolar}, and has purely {\em
nematic} order.  The Vicsek flock moves on average in the $\nhat$ direction,
while a nonequilibrium steady state with nematic order -- an active nematic --
cannot tell forward from back, and so does not drift on the average. The nature
of order in such active nematics is the subject of our study. Our main concern is whether 
the interplay of nonlinearity and fluctuations stiffens the order-parameter 
fluctuations in active nematics as it does \cite{tonertu} in polar ordered phases, 
leading to true long-range order in two dimensions.  

Here are our main results. (i) We elucidate the route to the equation of motion
for the nematic orientational order parameter, taking care to distinguish the
constraints introduced purely by rotation invariance, and hence applicable to
both active and equilibrium systems, from those arising specifically in the
thermal-equilibrium limit. (ii) We show that the two quadratic nonlinearities in the equation of motion have independent coefficients, unlike in the equilibrium case where they are determined by a single parameter. In both equilibrium and active nematics power-counting shows that the nonlinearities are marginal, but such analysis cannot distinguish
marginally relevant from marginally irrelevant. (iii) In a certain limit of
parameter values, our equation of motion can be mapped to the noisy
two-dimensional Burgers \cite{FNS} and KPZ \cite{BurgerKPZ} equations, but with a velocity
field ${\bf v}$ satisfying the peculiar condition $\partial_x
v_x-\partial_zv_z=0$, which is neither solenoidal nor irrotational. (iv) The similarity to the Burgers problem ends there: our dynamical renormalization-group treatment shows that the nonlinearities are marginally {\em irrelevant} in our theory, in the Burgers limit as well as in
general. Active nematics thus have only quasi-long-range order. Although
disappointing if one is looking for novelty in nonequilibrium systems, this
negative result reinforces the findings of a numerical study
\cite{chateginellimontagne} of an apolar generalization of the Vicsek model. 

This paper is organized as follows. In section \ref{eom} we construct the coarse-grained equations of motion for the nematic order parameter, highlighting the differences between equilibrium and active systems. In section \ref{burgers} we examine the relation of our equations of motion to the Burgers and KPZ equations, in a special high-symmetry limit. 
In section \ref{DRG} we outline the dynamic renormalization group (DRG) treatment with which we extract the long-time, long-wavelength properties of correlation functions in our system. Further calculational details are relegated to the Appendix. The paper closes in section \ref{discussion} with a discussion of possible future directions.

\section{Equation of motion}
\label{eom}
We now construct the equations of motion for an active nematic. Since we are considering a system that can undergo apolar orientational ordering, one of the slow variables 
for a coarse-grained description of the dynamics is the traceless symmetric second-rank tensor nematic order parameter $\Q$ \cite{degp}. The magnitude of $\Q$ is slow upon approach to the ordering transition, and the fluctuations of its principal axis are the broken-symmetry modes of the ordered phase.  If the system were isolated, mass and momentum would be conserved within the system and the corresponding densities $\rho$ and ${\bf J} = \rho {\bf v}$, 
${\bf v}$ being the velocity field, would be slow variables as well \cite{temperaturefootnote}. However, we will consider a system adsorbed on a solid surface which acts as a momentum sink, thus turning ${\bf J}$ or ${\bf v}$ into a fast variable, and allow deposition and evaporation \cite{khandkarbarma}, i.e., birth and death \cite{tonerpc}, thus rendering $\rho$ fast as well. We will start from a complete dynamical description, eliminate the fast $\rho$ and ${\bf J}$,  and obtain the dynamics of $\Q$ alone. 

For a system where particles can enter and leave the system in the bulk, the density obeys 
\begin{equation}
\frac{\partial \rho}{\partial t} =  -\gamma \rho + \beta -\nabla \cdot {\bf J} + f_{\rho}.
\label{eqnumgen}
\end{equation}
The third term on the right of (\ref{eqnumgen}) contains the number-conserving motion of particles on the substrate. The random adsorption and desorption of discrete particles has two effects. In the mean, conditioned on a given local density $\rho({\bf r},t)$, it leads to the $\gamma$ and $\beta$ terms. Fluctuations about this average effect lead to the nonconserving spatiotemporally white noise $f_{\rho}$. A steady, spatially uniform state has mean density $\rho_0 \equiv \beta/\gamma$.  Newton's second law for the momentum density $m{\bf J}$ reads 
\beq
\label{eqmomgen}
m \frac{\partial {\bf J}}{\partial t}  = - \Gamma {\bf v} + {\bf f}_{R} - \nabla \cdot \sigma 
\eeq
The first term on the right hand side of (\ref{eqmomgen}) is friction due to the substrate, with a kinetic coefficient $\Gamma$. The random agitation of the particles as a result of thermal motions, biochemical stochasticity, or dynamical chaos is modelled in the simplest possible manner by the spatiotemporally white Gaussian noise ${\bf f}_{R}$. This noise is nonconserving, i.e., its strength is nonvanishing at zero wavenumber, since the dynamics is not momentum-conserving. The last term contains all effects arising from interactions of the particles with each other, and thus takes the momentum-conserving form of the divergence of a stress tensor $\sigma$. In principle $\sigma$ contains stresses coming from the free-energy functional for $\Q$ (see below)\cite{qviscous}
These, however, are readily seen \cite{sradititoner} to be irrelevant at large lengthscales compared to the contribution $\sigma^a = w_1 \rho \Q$ coming from the active nature of the particles \cite{activestressrefs}. 
 

The equation of motion for the orientational order parameter $\Q$ including coupling to the velocity field \cite{dforster, pdolmsted} is 
\begin{equation}
\frac{\partial  \Q}{\partial t} + {\bf v} \cdot \nabla \Q  = \Gamma G + (\alpha_{0} {\bf \kappa} + \alpha_{1} {\bf \kappa}\cdot \Q)_{ST}  
                                                                              + \Omega \cdot \Q - \Q \cdot \Omega
\label{Qeom}
\end{equation}
where 
 ${\bf \kappa} = [\nabla {\bf v} + (\nabla {\bf v})^{T}]/2$ 
and 
$\Omega = [\nabla {\bf v} - (\nabla {\bf v})^{T}]/2$
are the shear rate and vorticity tensor respectively, 
$\Gamma$ is a kinetic coefficient \cite{kincoeffootnote}, and 
the parameters $\alpha_{0}$ and $\alpha_{1}$ characterise the coupling of orientation to flow. The molecular field 
$ G = -\delta F/\delta  \Q$ is obtained from an extended Landau-de Gennes free energy 
\begin{align}
F & = \int d^d x [{a \over 2} \mbox{Tr}\Q^2 + {u \over 4}  (\mbox{Tr}\Q^2)^2 + {K \over 2} 
(\nabla_i Q_{kl})^2  \notag \\ 
& + \bar{K} Q_{ij} \nabla_i Q_{kl} \nabla_j Q_{kl} + C Q_{ij} \nabla_i \nabla_j \rho] + \Phi[\rho] 
\label{ldgfree}
\end{align}
where we have left out terms cubic in $\Q$ as these vanish \cite{degp} in dimension $d=2$. 
The density $\rho$ enters $F$ through the functional $\Phi$, the quadrupolar coupling term with coefficient $C$, and the $\rho$-dependence of parameters in $f$.   
On timescales much larger than $1/\gamma$ and $m/\Gamma$,\ the density and momentum  equations (\ref{eqnumgen}) and (\ref{eqmomgen}) become constitutive relations determining $\rho$ and ${\bf J}$ in terms of the slow field $\Q$. 
Eq. (\ref{eqnumgen} tells us we can replace $\rho$ everywhere by $\rho_0$  to leading order in gradients, 
and (\ref{eqmomgen}) 
becomes 
\beq
\label{veff}
{\bf v} \simeq -{w_1 \rho_0 \over \Gamma} \nabla \cdot \Q
\eeq
apart from noise terms. 
The molecular field $G$ in (\ref{Qeom}) contains a term of the form $\Q \nabla \nabla \Q$, and one of the form $\nabla \Q \nabla \Q$, whose coefficients will be related as both terms arise as variational derivatives of the single $\bar{K}$ term in $F$ [Eq. (\ref{ldgfree}). Replacing ${\bf v}$ by its expression (\ref{veff}) in Eq. (\ref{Qeom}) will give rise to additional terms of that form, controlled by the activity parameter $w_1$. As a result,  the $\Q \nabla \nabla \Q$ and $\nabla \Q \nabla \Q$ terms in the effective equation of motion for $\Q$ cannot be combined into the variational derivative of a scalar functional, and will have two independent coefficients. We will explore below the consequences of the existence of two independent nonlinear couplings. In space dimension $d=2$ the order-parameter tensor has the simple form 
\beq
\label{Qdef}
 \Q = {S \over 2}\left(\begin{array}{cc}
 \cos 2\theta & \sin 2\theta\\
\sin 2\theta & -\cos 2\theta\end{array} \right),
\eeq
where the scalar order parameter $S$ measures the magnitude of nematic order and 
$\theta$ is the angle from a reference direction. Let us work in the nematic phase, where we can take $S=$ constant and define $\theta = 0$ along axis of mean macroscopic orientation. 
Eq. (\ref{veff}) for small $\theta$ becomes 
\begin{equation}
{\bf v} =  -\bar \Gamma^{-1}(\partial_{z}\theta, \partial_{x}\theta), 
\label{veffnem2d}
\end{equation}   
neither a gradient nor a curl, $\bar{\Gamma}$ being a constant determined by those in (\ref{eqnumgen}) - (\ref{Qdef}). Substituting ${\bf v}$ in (\ref{Qeom}) by its expression (\ref{veffnem2d}), writing $ \Q$ in terms of $\theta$ as in (\ref{Qdef}), treating $S$ as constant, and including noise terms,   
we obtain 
\begin{equation}
\label{Qeomeff}
\frac{\partial \theta}{\partial t} = A_{1}\partial_{x}^2 \theta + A_{2}\partial_{z}^2 \theta + \lambda_{1} \partial_{x}\theta \partial_{z}\theta + \lambda_{2} \theta \partial_{x} \partial_{z}\theta + f_{\theta}
\end{equation} 
to order $\theta^2$, where the additive\cite{multimplicativenoisefootnote}
non-conserving Gaussian white noise$f_{\theta}$ satisfies 
\begin{equation}
\label{Qnoise}
<f_{\theta}({\bf r}, t)f_{\theta}({\bf r^{'}}, t^{'})>
= 2D_{0} \delta({\bf r}-{\bf r^{'}})\delta(t-t^{'})
\end{equation}
with a noise strength $D_{0}$. All the coefficients in (\ref{Qeomeff}) and (\ref{Qnoise}) are related to those in (\ref{eqnumgen}) - (\ref{Qeom}), the corresponding noise strengths, and the scalar order parameter $S$. As a consequence of rotation invariance, i.e., the fact that the underlying equation of motion in terms of $\Q$ has a frame-independent form, we find 
\beq
\label{rotinv}
2(A_{1}-A_{2}) = \lambda_{2}. 
\eeq
It is therefore convenient to re-express them as 
\beq
A_{1} = A_{0}+\lambda_{2}/4; \qquad  A_{2} = A_{0}-\lambda_{2}/4  
\label{A0eqn}
\eeq
Without the detailed derivation above, it would have been hard to guess the form of the equations of motion and the constraints on the parameters. Note that $\lambda_1$ and $\lambda_2$ are in general independent, as we argued above. We will comment below on the relation they satisfy in the special case of an equilibrium nematic. Eqs. (\ref{Qeomeff}) and (\ref{rotinv}) can also be obtained from a microscopic model of collisional dynamics of apolar particles \cite{collisionalderivation}. 

\subsection{Equilibrium limit}
\label{eqmlimit}
 The energy cost of elastic deformations and, hence, the thermal equilibrium statistics of configurations, of a two-dimensional nematic are governed by the Frank free energy \cite{frankandors, degp, NP}
\begin{equation}
H=\int{[\frac{K_1}{2}(\nabla \cdot {\bf n})^2 + \frac{K_3}{2} (\nabla \times {\bf n})^2]d^2 r}, 
\label{frankdir2d}
\end{equation}  
a functional of the director field ${\bf n} = (\cos \theta, \sin \theta)$, with splay and bend elastic moduli $K_{1}$ and $K_{3}$. To cubic order in $\theta({\bf r})$ 
\begin{align} 
H /k_B T & = {A_{3} \over 2} \int{d_{2}\bf r}[[\partial_{x}\theta({\bf r})]^{2} + (1+\Delta)[\partial_{z}\theta({\bf r})]^{2}  \notag \\ 
           & -2\Delta  \theta({\bf r})[\partial_{x}\theta({\bf r})\partial_{z}\theta({\bf r})]]  
\label{franktheta2d}
\end{align}
where $A_3 = K_3/k_B T$ and $\Delta = \frac{(K_{1}-K_{3})}{K_{3}}$. The purely relaxational dynamics of the angle field $\theta$, at thermal equilibrium consistent with (\ref{franktheta2d}), reads
\begin{align}
\frac{\partial \theta}{\partial t} & = A_{3}\partial_{x}^2 \theta + (1 + \Delta)A_3\partial_{z}^2 \theta + \lambda_{1} \partial_{x}\theta \partial_{z}\theta  \notag \\
&+ \lambda_{2} \theta \partial_{x} \partial_{z}\theta + f_{\theta}
\label{equilthetadyn}
\end{align} 
where $\langle f_{\theta}({\bf r}, t) f_{\theta}({\bf 0}, 0) \rangle = 2 \delta({\bf r}) \delta(t)$, and a kinetic coefficient has been absorbed into a time-rescaling. 
The nonlinearities in  (\ref{equilthetadyn}) have the same form as in (\ref{Qeomeff}), but the couplings are not independent: $2\lambda_1= \lambda_2 = -2A_3$, since both come from the same anharmonic term in the free energy (\ref{franktheta2d}). In addition, the nonlinearity is connected  to the diffusion anisotropy: $2[A_3 - (1 + \Delta)A_3] = \lambda_2$ as required by rotation invariance. Eq. (\ref{equilthetadyn}) is simply the limit $2\lambda_1= \lambda_2$ of (\ref{Qeomeff}). 

A static renormalization-group treatment of the $2d$ equilibrium nematic \cite{NP} with Hamiltonian 
(\ref{franktheta2d}) showed that $\Delta$ was marginally {\em ir}relevant, and that the large-scale behaviour of the system was governed by a fixed point with $\Delta = 0$, i.e., a single, finite Frank constant for both splay and bend. The dynamics of the active nematic does not correspond to downhill motion with respect to a free-energy functional, and the two nonlinear terms thus have independent coefficients. Their (marginal) relevance or otherwise must be established by a dynamic renormalization-group study of the equation of motion (\ref{Qeomeff}), which we present in section \ref{DRG}. 

\subsection{Burgers  equation} 
\label{burgers} 
The structure of (\ref{Qeomeff}) in a certain special limit merits some attention. If we switch off the $\lambda_2$-nonlinearity, equation (\ref{Qeomeff}) has a higher symmetry than in general, \textit{viz.}, under $\theta \rightarrow \theta + \mbox{constt}$ without a corresponding transformation of the coordinates. In addition, it is invariant under $x \leftrightarrow z$, which allows us to rescale the equations so that the diffusion of $\theta$ is isotropic: 
\beq
\label{kpzlike}
\frac{\partial \theta}{\partial t} = A{\bf \nabla}^{2} \theta + \lambda \partial_{x}\theta \partial_{z}\theta + f_{\theta}
\eeq 
with a spatiotemporally white noise $f_{\theta}$  as in  (\ref{Qnoise}).
This equation for $\lambda \neq 0$ cannot correspond to an equilibrium system, because the sole surviving  nonlinear term  
$\lambda \partial_{x}\theta \partial_{z}\theta$ cannot be written as $\delta A / \delta \theta({\bf x})$ for any scalar functional $A[\theta]$\cite{noneqmtermfootnote}
Note the similarity of (\ref{kpzlike}) to the KPZ equation \cite{BurgerKPZ} for the height field of a driven interface. Extending the analogy, it is easy to see that the velocity field 
${\bf v} = (\partial_{z}\theta, \partial_{x}\theta)$ 
as in  (\ref{veffnem2d}) obeys the Burgers-like equation \cite{BurgerKPZ,FNS}
\begin{equation}
\label{burgerslike}
\frac{\partial {\bf v}}{\partial t} = A\nabla^{2}{\bf v} + \lambda ({\bf v} \cdot \nabla) {\bf v} + {\bf f}_{{\bf v}}
\end{equation}
with a conserving noise ${\bf f}_{{\bf v}} = (\partial_{z} f_{\theta}, \partial_{x}f_{\theta})$. 
The curl-free condition of a traditional Burgers velocity field is replaced in our case by $\partial_{x}v_{x} - \partial_{z}v_{z} = 0$, which amounts to equal extension rates along $x$ and $z$. In the $2d$ randomly-forced Burgers-KPZ problem, the nonlinearity is known \cite{FNS, BurgerKPZ} 
to be marginally relevant, so that the large-scale long-time behaviour is governed by a strong-coupling fixed point inaccessible to a perturbative RG. It is natural to ask what happens in the seemingly similar problem at hand. 
\subsubsection{Galilean invariance} 
\label{galinv}
Eqns. (\ref{kpzlike}) and (\ref{burgerslike}) are invariant under the infinitesimal Galilean boost 
\beq 
\label{galboost}
{\bf x} \to {\bf x} - {\bf u}t
\eeq
\beq 
\label{galtheta}
\theta \to \theta + \tilde{\bf u} \cdot {\bf x}
\eeq
or equivalently 
\beq
\label{galv}
{\bf v} \to {\bf v} + {\bf u}
\eeq
where 
\beq
\label{equ}
\tilde{\bf u} = (u_z, u_x)
\eeq
inverts the vector components of ${\bf u}$. 
By analogy to the results of 
\cite{FNS} and \cite{BurgerKPZ} this invariance implies that the nonlinear-coupling $\lambda$
does not renormalise in this special limit.

\section{Renormalization group theory}
\label{DRG}
In this section we outline our one-loop dynamic renormalization group 
(DRG) analysis of the large-scale, long-time behaviour of Eq. (\ref{Qeomeff}). Our treatment is general, allowing for two independent coupling strengths $\lambda_1$, $\lambda_2$, but we will examine the $\lambda_2 \to 0$ limit of section \ref{burgers} as well. 
We present only the key steps of the calculation, relegating details to the Appendices.

The momentum-shell dynamical renormalization group (DRG) \cite{SMa, SMab, HH, FNS} 
consists of two steps. Consider a system with physical fields described by Fourier modes with wavevector ${\bf q}$ with $0 \leq q \equiv |{\bf q}| < \Lambda$, the ultraviolet (UV) cutoff. First: eliminate modes with $\Lambda e^{-l} \leq q < \Lambda$, 
by solving for them in terms of those in $0 \leq q < \Lambda e^{-l}$ and the noise, and average over that part of the noise whose wavenumber lies in $[\Lambda e^{-l}, \, \Lambda)$.  Second: rescale space, time, and dynamical variables to restore the cutoff $\Lambda$ and to preserve the form of the equations of motion to the extent possible. The result is an equation of motion in which the parameters have changed from their initial values, call them $\{K_0\}$, to $l$-dependent values $\{K(l)\}$. Now, correlation functions at small wavenumber can be calculated either from the original equations of motion or from those obtained after the above two steps. This key observation leads to a homogeneity relation between correlation functions 
\beq
\label{homog}
C({\bf q}, \omega; \{K_0\}) = e^{fl}C({\bf q}e^l, \omega e^{zl}; \{K(l)\}).  
\eeq
that can be used to calculate long-wavelength correlations with particular ease if the couplings flow to a small fixed-point value $\{K(\infty)\}$ under iteration of the above transformation. 
Let us carry out this process for our model, Eq. (\ref{Qeomeff}). 

We insert the decomposition \cite{intkomagefootnote}
$\theta({\bf r}, t) = \int_{q<\Lambda, \omega} \theta({\bf q}, \omega) \exp{(i{\bf q}\cdot {\bf r} - i\omega t)}$ 
into (\ref{Qeomeff}) to obtain the $\theta$ equation in Fourier space: 
\begin{align}
\theta({\bf q}, \omega) & = G_{0}({\bf q}, \omega)f_{\theta}({\bf q}, \omega)  -  G_{0}({\bf q}, \omega) \notag \\
&   \int_{k\Omega}M({\bf k}, {\bf q}-{\bf k}){\theta({\bf k}, \Omega) \theta({\bf q} - {\bf k}, \omega - \Omega)} 
\label{Qeomft}
\end{align}
where 
\begin{equation}
G_{0}({\bf q}, \omega) = [-i \omega + A_{1}q_{x}^2 + A_{2}q_{z}^2]^{-1}  
\label{Qpropagft}
\end{equation}
is the bare propagator,  
\begin{align}
M({\bf k}, {\bf q}-{\bf k}) & = \frac{\lambda_{1}}{2}[k_{x}(q_{z}-k_{z}) + k_{z}(q_{x}-k_{x})] \notag  \\
& + \frac{\lambda_{2}}{2}[k_{x}k_{z} + (q_{x}-k_{x})(q_{z}-k_{z})] 
\label{Qvtxft}
\end{align}
the bare vertex, and the Fourier transform $f_{\theta}({\bf q}, \omega)$ of the Gaussian spatiotemporally white noise in (\ref{Qeomeff}) has autocorrelation 
\begin{equation}
\langle f_{\theta}({\bf q}, \omega) f_{\theta}({\bf q}^{'}, \omega^{'})\rangle  = 
2D_{0}(2\pi)^{2+1} 
                                                                  \delta({\bf q}+{\bf q}^{'})\delta(\omega+\omega^{'})
\label{Qnoisecorrft}
\end{equation}
\begin{figure}[htbp]
  \begin{center}
       \subfigure{(a)\includegraphics[height=2.3cm, width=7.0cm]{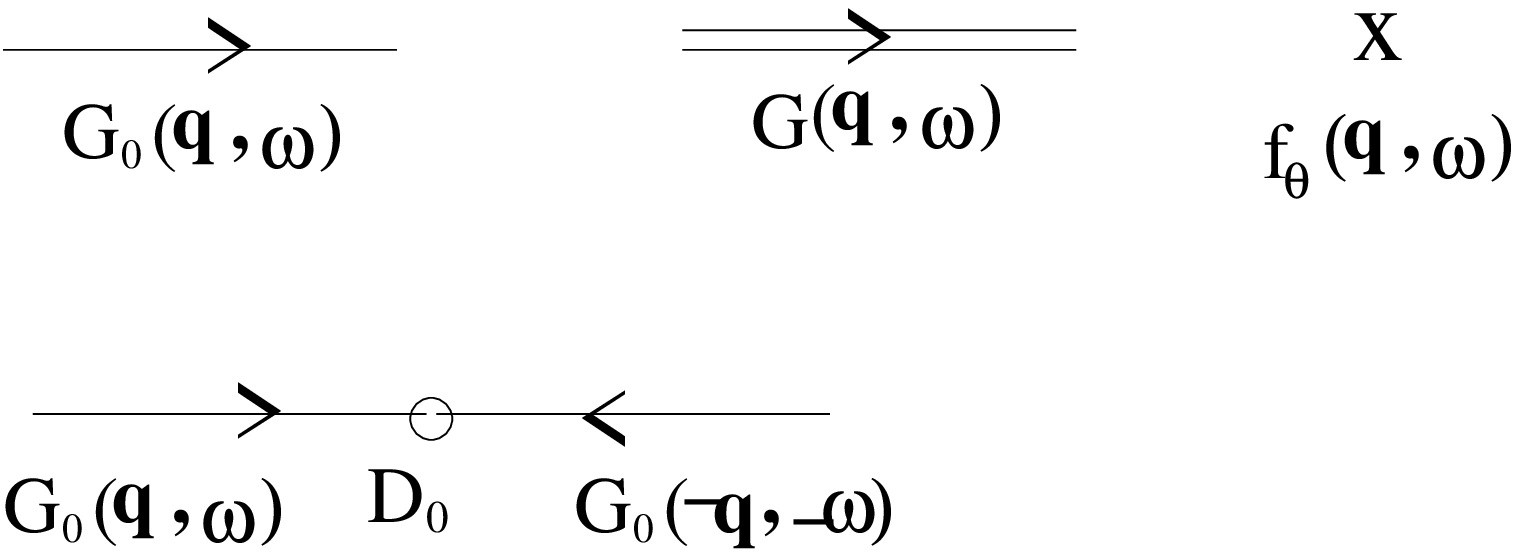}}
      \subfigure{(b)\includegraphics[height=2.3cm, width=7.0cm]{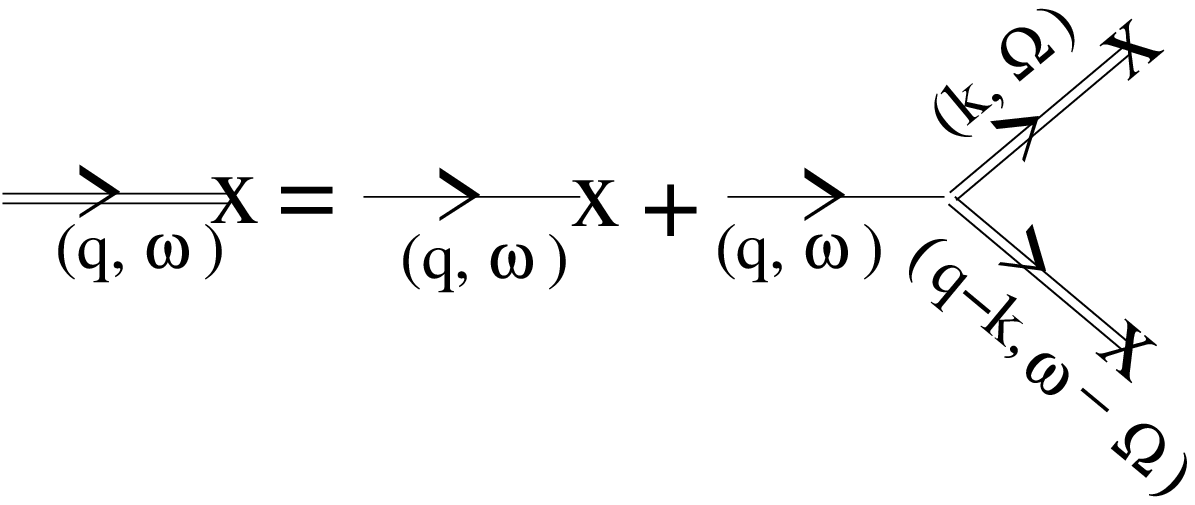}}
    \caption{(a) Definition of symbols. (b) Diagram for full non-linear equation (\ref{Qeomft}) 
in Fourier space. The left hand side of the pictorial equation is the full  solution to
$\theta( {\bf q}, \omega) = G({\bf q}, \omega) f_{\theta}({\bf q}, \omega)$,
where $ G({\bf q}, \omega)$ is the full propagator.
The first part on the right hand side is the zeroth order solution to (\ref{Qeomft})
$\theta({\bf q}, \omega) = G_{0}({\bf q}, \omega) f_{\theta}({\bf q}, \omega)$
and the second term  is the contribution of the nonlinearity.}
    \label{symbolsfig}
  \end{center}
\end{figure}

\begin{figure}[htbp]
  \begin{center}
      \subfigure(a){\includegraphics[height=2.1cm, width=7.0cm]{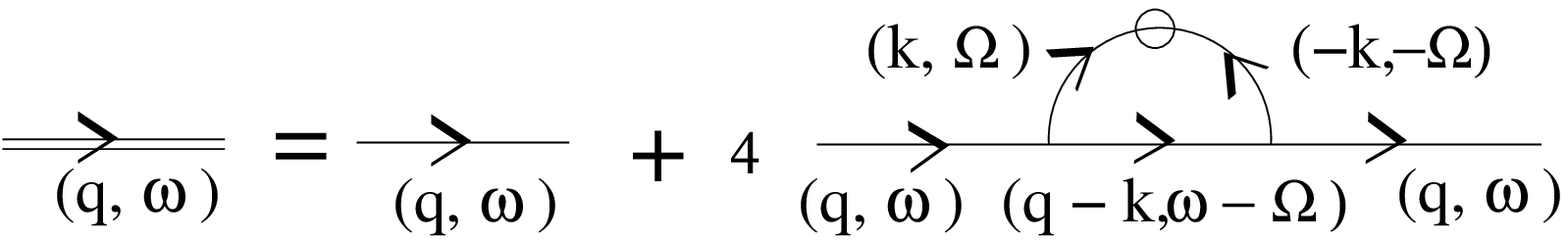}}
       \subfigure(b){\includegraphics[height=2.1cm, width=7.0cm]{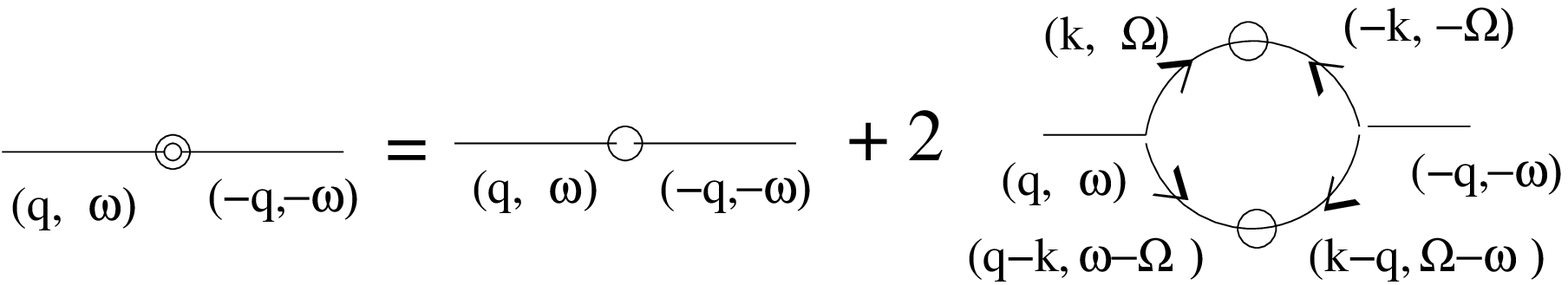}}  
       \subfigure(c){\includegraphics[height=2.1cm, width=7.0cm]{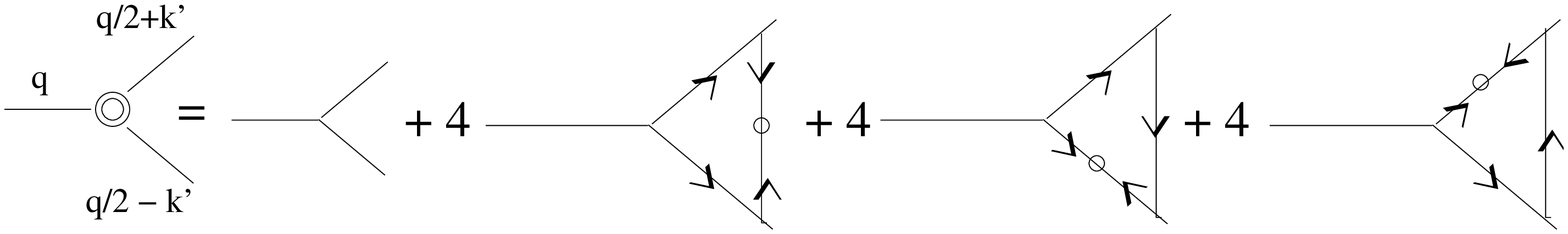}} 
    \caption{(a) Graph for propagator $G({\bf q}, \omega)$. The left hand side with a double line is the full propagator, the first term on the right hand side is the zeroth order and the second term is the one-loop correction. (b) Graph for force density $D({\bf q}, \omega)$ defined by  (\ref{Qnoisecorrft}). The second term on the right hand side is the one-loop correction. (c) Graph for the three-point vertex function. The structure with three legs with one incoming and 
two outgoing is the vertex $-\frac{1}{(2\pi)^{2+1}}\int{M({\bf k}, {\bf q}-{\bf k})}$. The three graphs are $\Gamma_{a}$, $\Gamma_{b}$ and $\Gamma_{c}$.}
\label{perttheoryfig}
\end{center}
\end{figure}

Eq. (\ref{Qeomft}) can be represented graphically as in Fig. \ref{symbolsfig}. 
A perturbative approach to solving (\ref{Qeomft})  generates corrections that can be expressed in terms of Feynman graphs of three types --  propagator, noise strength  and nonlinearities -- given in Fig. \ref{perttheoryfig}. 
\subsection{Propagator calculation}
\label{propagatorcal}
The effective propagator  $G({\bf q}, \omega)$ 
[defined by $\theta({\bf q}, \omega) \equiv G({\bf q}, \omega) f_{\theta}({\bf q}, \omega)$] 
is given perturbatively in Fig. \ref{perttheoryfig}(a). 
The averaging over the noise is performed using  (\ref{Qnoisecorrft}). 
The one-loop correction to the propagator is  
\begin{align}
 G({\bf q}, \omega) & =  G_{0}({\bf q}, \omega) + 4 G_{0}^{2}(({\bf q}, \omega)\times 2 D_{0} \notag \\
                     &    \int_{k\Omega}M({\bf k}, {\bf q} - {\bf k}) M(-{\bf k}, {\bf q})G_{0}({\bf k}, \Omega) \notag \\ 
                     &    G_{0}(-{\bf k}, -\Omega) G_{0}({\bf q} - {\bf k}, \omega - \Omega)
\label{propagatorft}
\end{align}      
or 
\begin{equation}
G^{-1}({\bf q}, \omega) =  G_{0}^{-1}({\bf q}, \omega) - \Sigma ({\bf q}, \omega)
\label{fullpropagator}
\end{equation}
with a self-energy  
\begin{align}
\Sigma ({\bf q}, \omega) & = 4 \times 2D_{0} \int_{k\Omega}M({\bf k}, {\bf q} - {\bf k})M(-{\bf k}, {\bf q})  \notag \\
                         &   G_{0}({\bf k}, \Omega) G_{0}(-{\bf k}, -\Omega)G_{0}({\bf q} - {\bf k}, \omega - \Omega)    
\label{selfenergy}
\end{align}      
 where the combinatorial factor of four represents possible 
noise contractions leading to Fig \ref{perttheoryfig} (a). 
A few steps of calculation of the integrals are performed in Appendix \ref{appA}. 
For small wavenumber ${\bf q}$ and for $\omega \rightarrow 0$, the result of integrating out a shell between $\Lambda e^{-l}$ and $\Lambda$ in ${\bf q}$ space is the self-energy.  
\begin{align}
\Sigma ({\bf q}, 0) & = \frac{l}{4\pi}\bigg [-\frac{G_{2}(\bar \lambda_{1}, \bar \lambda_{2})}{8}(A_{1}q_{x}^{2} + A_{2}q_{y}^{2}) \notag \\
                         & + \frac{G_{3}(\bar \lambda_{1}, \bar \lambda_{2}) A_{1}A_{2}}{(\sqrt{A_{1}} + \sqrt{A_{1}})^{2}} \bigg]
\label{eqselfenergy}
\end{align}   
where
\begin{align}
&  G_{2}(\bar \lambda_{1}, \bar \lambda_{2}) = (2\bar \lambda_{1}^{2} + \bar \lambda_{2}^{2} - 3\bar \lambda_{1}\bar \lambda_{2}) \notag \\
&  G_{3}(\bar \lambda_{1}, \bar \lambda_{2}) = (\bar \lambda_{2}^{2} -  \bar \lambda_{1}\bar \lambda_{2})    
\label{selfenergysymbols}
\end{align}
The dimensionless quantities $\bar \lambda_{1}$ and $\bar \lambda_{2}$ are defined by
\begin{equation}
 \bar \lambda_{i} \bar \lambda_{j} = \frac{\lambda_{i} \lambda_{j} D_{0}}{(A_{1}A_{2})^{3/2}}, \qquad i, j=1, 2
\label{lambdabars}
\end{equation}
When we implement the dynamical renormalization group, terms of order $q^{2}$ 
and of order 1 are generated though the self-energy. Terms of order $q^{2}$ will give corrections to the diffusion constants $(A_{1}, A_{2})$. 
What about the terms \cite{comparisonburgerfootnose}
of order 1, which also arise in  the analysis of Pelcovits {\em et al.} 
\cite{NP}?
As in \cite{NP}, we proceed by first ignoring the terms of order 1, whose coefficient is proportional
to one nonlinear coupling $\lambda_2$, and then, post facto, realise they too are (marginally) irrelevant because $\lambda_2$ itself is found to be marginally irrelevant.
Proceeding in this manner we find
\begin{align}
G^{-1}({\bf q}, 0) & = G_{0}^{-1}({\bf q}, 0) - \Sigma ({\bf q}, 0) \notag \\
                    & \sim \tilde{A_{1}}q_{x}^{2} + \tilde{A_{2}}q_{z}^{2}  \notag \\
                    & = A_{1}q_{x}^{2} + A_{2}q_{z}^{2}  \notag \\ 
                    & + \frac{G_{2}(\bar \lambda_{1}, \bar \lambda_{2})(A_{1}q_{x}^{2} + A_{2}q_{z}^{2}) l}{4\times 8\pi}  
\label{propagatorint}
\end{align}
That is,
\begin{align}
 & \tilde{A_{1}} = A_{1}[1+\frac{G_{2}(\bar \lambda_{1}, \bar \lambda_{2}) l}{4\times 8\pi}]; \notag  \\
 & \tilde{A_{2}} = A_{2}[1+\frac{G_{2}(\bar \lambda_{1}, \bar \lambda_{2}) l}{4\times 8\pi}].
\label{eq_anisodiff_interm}
\end{align}
These are the intermediate (one-loop graphical) corrections for anisotropic diffusion constants.  
\subsection{Vertex calculation}
\label{vertexcal}
From the full equation (\ref{Qeomft}) and (Fig \ref{symbolsfig}), 
the diagrams  contributing to the vertex correction are shown in (Fig \ref{perttheoryfig}(b)). 
There will be three types of diagrams, all with multiplicity 4, 
denoted by $\Gamma_{a}$, $\Gamma_{b}$ and $\Gamma_{c}$. 
The details of the calculation are given in Appendix \ref{appB}. 
The full vertex is defined as a combination of 
$\lambda_{1}$ and $\lambda_{2}$ equation (\ref{Qvtxft}). 
We study how this vertex evolves under the DRG and at the 
end of the calculation we can separate terms corresponding to $\lambda_{1}$ and $\lambda_{2}$. 
From (Fig \ref{perttheoryfig}(b)), expression for
\begin{align}
\Gamma_{a}({\bf q}, {\bf k_{1}}) & = 4 \times 2D_{0} \int_{k\Omega}M({\bf k}, {\bf q}- {\bf k}) \notag \\ 
                                 & \times  M(\frac{{\bf q}}{2} + {\bf k}_{1}, {\bf k} -\frac{{\bf q}}{2} - {\bf k}_{1}) \notag \\
                                 & \times  M(\frac{{\bf q}}{2} -  {\bf k_{1}}, -{\bf k} + \frac{{\bf q}}{2}  + {\bf k_{1}})  \notag \\
                                 & \times  \bigg|G_{0}({\bf k} - \frac{{\bf q}}{2} - {\bf k}_{1}, \Omega - \frac{\omega}{2}-\Omega_{1})\bigg|^{2} \notag \\
                                 & \times  G_{0}({\bf k}, \Omega) G_{0}({\bf q} - {\bf k}, \omega-\Omega)  
\label{eqGamma_a}
\end{align}
The integral as usual is over $\Lambda e^{-l} < q < \Lambda$. Similarly one can get expressions  for $\Gamma_{b}({\bf q}, {\bf k_{1}})$ and $\Gamma_{c}({\bf q}, {\bf k_{1}})$. Hence, adding contributions to all diagrams for the vertex, $ \Gamma_{a}({\bf q}, {\bf k_{1}})+\Gamma_{b}({\bf q}, {\bf k_{1}})+\Gamma_{c}({\bf q}, {\bf k_{1}})$ we can get the graphical corrections to the couplings $\lambda_{1}$ and $\lambda_{2}$. After a calculation as in Appendix \ref{appB}, the graphical corrections to  $\lambda_{1}$ and $\lambda_{2}$ are
\begin{align}
 & \tilde{\lambda}_{1} = \lambda_{1}[1 - \frac{F_{1}(\bar \lambda_{1}, \bar \lambda_{2})l}{2\times 8\pi}] \notag \\
 & \tilde{\lambda}_{2} = \lambda_{2}[1 - \frac{F_{2}(\bar \lambda_{1}, \bar \lambda_{2})l}{2\times 8\pi}] 
\label{lambda_interm}
\end{align}
$ F_{1}(\bar \lambda_{1}, \bar \lambda_{2})$, $F_{2}(\bar \lambda_{1}, \bar \lambda_{2})$ defined by, 
\begin{align}
 & F_{1}(\bar \lambda_{1}, \bar \lambda_{2}) = -2\bar \lambda_{1}\bar \lambda_{2} + 3 \bar \lambda_{2}^{2} + \bar \lambda_{2}^{3}/\bar \lambda_1 \notag \\ 
& F_{2}(\bar \lambda_{1}, \bar \lambda_{2}) = -4\bar \lambda_{2}\bar \lambda_{1} + 6\bar \lambda_{2}^{2}  
\label{Flambda_interm}
\end{align}
Note from  (\ref{Flambda_interm}) that  $F_{1}(\bar \lambda_{1}, 
\bar \lambda_{2}) = 0$, if $\lambda_{2}$ is zero. 
This says that there is no graphical correction to 
$\lambda_{1}$ if $\lambda_{2}$ is zero.
This is a result of the Galilean invariance in this limit, as pointed out in section \ref{galinv}.

\subsection{Noise strength renormalization}
\label{noiserenorm}
An effective noise strength $\tilde{D}$ can be defined by 
\begin{equation}
\langle \theta^{*}({\bf q}, \omega) \theta({\bf q}, \omega) \rangle = 2\tilde{D} G({\bf q}, \omega)G(-{\bf q}, -\omega).
\label{effnoisestrength}
\end{equation}
This quantity is calculated perturbatively by the series shown in 
(Fig {\ref{perttheoryfig}(c)}). To one-loop order 
\begin{align}
2\tilde{D} & = 2D_{0} + 2 (2D_{0})^{2} \int_{k\Omega} M({\bf k}, {\bf q}- {\bf k}) M(-{\bf k}, {\bf k}- {\bf q})  \notag \\ 
           & \times \bigg|G_{0}({\bf k}, \Omega)\bigg|^{2} \bigg| G_{0}({\bf q} - {\bf k}, \omega-\Omega)\bigg|^{2} 
\label{eq_D_oneloop}
\end{align} 
The integral in equation (\ref{eq_D_oneloop}) is performed in Appendix \ref{appC}. After doing the integrals, the graphical correction to $D_{0}$ is 
\begin{equation}
\tilde{D} =  D_{0}\bigg[ 1 + \frac{(\bar \lambda_{2} - \bar \lambda_{1})^{2} l}{2\times 8\pi}\bigg]
\label{eq_D_graph}
\end{equation}      
\subsubsection{The detailed balance limit}
\label{dblimit}
From equations (\ref{eq_anisodiff_interm}) and  (\ref{eq_D_graph}),  for $\lambda_{2}=0$ ($A_{1}=A_{2}=A$, $G_{2}(\bar \lambda_{1}, \bar \lambda_{2}) = (2\bar \lambda_{1}^{2} + \bar \lambda_{2}^{2} - 3\bar \lambda_{1}\bar \lambda_{2}) = 2\bar \lambda_{1}^{2}$ and $(\bar \lambda_{2} - \bar \lambda_{1})^{2} = \bar \lambda_{1}^{2}$), 
 i.e., $A$ and $D$ have the same graphical corrections. 
This suggests that detailed balance should obtain in the limit $\lambda_2=0$.
To discover this detailed balance let us write the  Fokker-Planck equation \cite{FP} for the probability distribution functional $P[\theta, t]$ of the $\theta$-field: 
\begin{align}
& \frac{\partial P}{\partial t} + \sum_{{\bf q}} \frac{\partial}{\partial \theta_{{\bf q}}} \bigg[D_{0} \frac{\partial}{\partial \theta_{-{\bf q}}} + A {\bf q}^{2} \theta_{{\bf q}} \notag \\ 
& + \frac{\lambda_{1}}{\sqrt{\Omega}}\sum_{{\bf l},{\bf m}} M({\bf l}, {\bf m}) \theta_{{\bf l}} \theta_{{\bf m}} \delta_{{\bf q}, {\bf l}+{\bf m}}\bigg]P = 0.  
\label{eq_theta_FP}
\end{align}
We guess that a Gaussian  probability distribution function 
\begin{equation}
P_{st} = N \exp \bigg[-\frac{1}{2}\sum_{{\bf q}} \frac{\theta_{{\bf q}}\theta_{-{\bf q}}}{<\theta_{{\bf q}}\theta_{-{\bf q}}>}\bigg]
\label{eq_gaussian_guess}
\end{equation} 
is a steady solution to equation (\ref{eq_theta_FP}),  $M({\bf l}, {\bf m}) = (l_{x}m_{y} + m_{x}l_{y})$, $N$ is a normalization factor and the two-point function  $<\theta_{{\bf q}}\theta_{-{\bf q}}> = (D_{0}/A)q^{-2}$. If this is so, the last term on the right of equation (\ref{eq_theta_FP}) should vanish if $P_{st}$ from equation (\ref{eq_gaussian_guess}) is inserted for $P$. Let us check this:
\begin{align}
& \bigg[\sum_{q,l,m}\frac{\partial}{\partial \theta_{q}}M({\bf l}, {\bf m})\theta_{{\bf l}} \theta_{{\bf m}} \delta_{{\bf q}, {\bf l}+{\bf m}}\bigg]P_{0} \notag \\
& = \sum_{q,l,m}M({\bf l}, {\bf m})\theta_{{\bf l}} \theta_{{\bf m}} \delta_{{\bf q}, {\bf l}+{\bf m}} \frac{\partial P_{0}}{\partial \theta_{q}} \notag \\ 
& = -P_{0}\frac{D_{0}}{A}\sum_{{\bf q},{\bf l},{\bf m}} {\bf q}^{2}M({\bf l}, {\bf m})\theta_{{\bf l}} \theta_{{\bf m}} \theta_{-{\bf q}}\delta_{{\bf q}, {\bf l}+{\bf m}}   
\label{eq_dbsum}
\end{align}
Using the symmetry $-{\bf q} \rightleftharpoons {\bf l} \rightleftharpoons {\bf m}$ in (\ref{eq_dbsum}) we get
\begin{align}
& \sum_{q,l,m} {\bf q}^{2}M({\bf l}, {\bf m})\theta_{{\bf l}} \theta_{{\bf m}} \theta _{-{\bf q}}\delta_{q, l+m} \notag \\
& = \frac{1}{3} \sum_{l,m} [M({\bf l}, {\bf m}) ({\bf l} + {\bf m})^2 + {\bf l}^2 M(-{\bf m}, {\bf l}+ {\bf m})  \notag \\
& + {\bf m}^2 M(-{\bf l}, {\bf l}+ {\bf m})]\theta_{{\bf l}} \theta_{{\bf m}} \theta _{-{\bf l}-{\bf m}}  
\label{eq_dbcancel}
\end{align}
The summation inside the square bracket in  (\ref{eq_dbcancel}) is zero. This means that for $\lambda_2=0$ the Gaussian defined in  (\ref{eq_gaussian_guess}), is a steady solution of the FP equation (\ref{eq_theta_FP}), consistent with the detailed balance noted after equation (\ref{eq_D_graph}) in this limit. In particular, we can already conclude that there is no singular renormalization of the stiffnesses in the Burgers-like limit of the model, as the equal-time correlators of $\theta$ can be obtained directly from the Gaussian probability distribution function (\ref{eq_gaussian_guess}). 
\subsection{Full RG Analysis}
\label{rganalysis}
We now return to the general case $\lambda_1$, $\lambda_2$ nonzero. Substituting results from (\ref{eq_anisodiff_interm}), (\ref{lambda_interm}) and  (\ref{eq_D_graph}) to  (\ref{Qeomft}), gives the intermediate equation for 
$\theta^<({\bf q}, \omega) $ (without rescaling) 
\begin{align}
 \theta^{<}_{l}({\bf q}, \omega) & = G_{l}({\bf q}, \omega)(f_{l \theta}({\bf q}, \omega) + \Delta f_{\theta}({\bf q}, \omega)) \notag \\ 
                                 & - G_{l}({\bf q}, \omega)\int_{k\Omega}M_{l}({\bf k}, {\bf q}-{\bf k}) \notag \\ 
                                 & \times \theta^{<}({\bf k}, \Omega)\theta^{<}({\bf q} - {\bf k}, \omega - \Omega), 
\label{fullthetaless}
\end{align}
where the propagator at this intermediate stage is 
\begin{equation}
 G_{l}({\bf q}, \omega) = (-i\omega + \tilde{A}_{1}q_{x}^{2}+\tilde{A}_{2}q_{z}^{2})^{-1},
\label{intempropagator}
\end{equation}
with $\tilde{A}_{1}$ and $\tilde{A}_{2}$ given by  (\ref{eq_anisodiff_interm}) and $0< |{\bf q}| < \Lambda e^{-l}$, unlike the original equation, which is defined on the large range $0<|{\bf q}| < \Lambda $.\\
Next rescale variables to preserve the form of the 
original equation: 
\begin{align}
& q^{'} = qe^{l}; \qquad 
\omega^{'} = \omega e^{\alpha(l)}; \qquad  \notag \\
& \theta^{<}({\bf q}, \omega) = \xi(l) \theta^{'}({\bf q}^{'}, \omega^{'}). 
\label{primedvariable}
\end{align}
Thus the new variable ${\bf q}'$ is defined on the same 
interval $0<|{\bf q}'| < \Lambda $ as the wave-vector ${\bf q}$ 
in the original equation. In terms of the new variables, 
the intermediate equation for 
$\theta'({\bf q}', \omega')$ is 
\begin{align} 
 \theta^{'}({\bf q}^{'}, \omega^{'}) & = G(l)({\bf q}^{'}, \omega^{'})f^{'}_{\theta}({\bf q}^{'}, \omega^{'}) \notag \\
                                     & -  G(l)({\bf q}^{'}, \omega^{'})\int_{k^{'}\Omega^{'}}M(l)({\bf k}^{'}, {\bf q}^{'}-{\bf k}^{'}) \notag  \\  
                                     & \times  \theta^{'}({\bf k}^{'}, \Omega^{'}) \theta^{'}({\bf q}^{'} - {\bf k}^{'}, \omega^{'} - \Omega^{'}), 
\label{thetaprime}
\end{align}
where 
\begin{equation}
 G(l)({\bf q}^{'}, \omega^{'}) = [-i\omega + A_{1}(l)q_{x}^{'2} + A_{2}(l)q_{z}^{'2}]^{-1}
\label{propagatorprime}
\end{equation}
with 
\begin{equation}
 A_{1}(l) = \tilde{A}_{1} e^{\alpha(l) - 2l}; \qquad  
A_{2}(l) = \tilde{A}_{2} e^{\alpha(l) - 2l};
\label{Ainterm}
\end{equation}
\begin{equation}
f^{'}_{\theta}({\bf q}^{'}, \omega^{'}) = f^{<}_{\theta}({\bf q}, \omega) e^{\alpha(l)} \xi^{-1}(l) 
\label{noiseprime}
\end{equation}  
\begin{align} 
 M(l)({\bf k}^{'}, {\bf q}^{'}-{\bf k}^{'}) &= \frac{\lambda_{1}(l)}{2}\bigg [k_{x}^{'}(q_{z}^{'}-k_{z}^{'})  \notag \\ 
                                             & + k_{z}^{'}(q_{x}^{'}-k_{x}^{'})\bigg ] + \frac{\lambda_{2}(l)}{2} \notag \\
                                             &  \times \bigg[k_{x}^{'}k_{z}^{'} + (q_{x}^{'}-k_{x}^{'})(q_{z}^{'}-k_{z}^{'})\bigg]  
\label{vertexprime}
\end{align}
where $\lambda_{1}(l)$ and $\lambda_{2}(l)$ are rescaled nonlinearities given by
\begin{equation}
\lambda_{1}(l) =  \tilde{\lambda}_{1} \xi(l) e^{-(d+2)l}; \qquad
\lambda_{2}(l) =  \tilde{\lambda}_{2} \xi(l) e^{-(d+2)l}
\label{lambdainterm}
\end{equation}
The correlation function characterising the force $f^{'}_{\theta}({\bf q}^{'}, \omega^{'}) $, given by expression  (\ref{noiseprime}), can be constructed using definition (\ref{Qnoisecorrft}) and the new set of variables (\ref{primedvariable})
\begin{align}
<f_{\theta}^{'}({\bf q}, \omega) f_{\theta}^{'}({\bf q}^{'}, \omega^{'})>   = & 
2D(l)(2\pi)^{2+1} 
\delta({\bf q}+{\bf q}^{'}) \notag \\ 
 & \delta(\omega+\omega^{'})
\label{Qnoisecorrftprime}
\end{align}
with
\begin{equation}
D(l) = \tilde{D}e^{(3\alpha(l) + dl)}\xi^{-2}(l)
\label{forcedl}
\end{equation}
where $d=2$ and all tilde variables correspond to the graphically corrected quantities in  (\ref{eq_anisodiff_interm}), (\ref{lambda_interm}) and  (\ref{eq_D_graph}). 
Substituting for the expressions for all tilde variables 
\begin{align}
 & A_{1}(l) = A_{1}[1+\frac{l G_{2}(\bar \lambda_{1}, \bar \lambda_{2})}{4\times 8\pi}]e^{\alpha(l)-2l},  \notag \\
 & A_{2}(l) = A_{2}[1+\frac{l G_{2}(\bar \lambda_{1}, \bar \lambda_{2})}{4\times 8\pi}]e^{\alpha(l)-2l}, \notag \\
 & \lambda_{1}(l) = \lambda_{1}[1 - \frac{l F_{1}(\bar \lambda_{1}, \bar \lambda_{2})}{4 \times 8\pi}]e^{-4l}\xi(l), \notag \\
 & \lambda_{2}(l) = \lambda_{2}[1 - \frac{l F_{2}(\bar \lambda_{1}, \bar \lambda_{2})}{4 \times 8\pi}]e^{-4l}\xi(l), \notag \\
 & D(l) = D[1+ \frac{l (\bar \lambda_{2} - \bar \lambda_{1})^{2}}{2 \times 8\pi}] e^{3\alpha(l)+2l}\xi^{-2}(l).  
\label{intermvariables}
\end{align}
\subsection{Recursion relation}
\label{recursionrelation} 
\begin{figure}[htbp]
  \begin{center}
      {\includegraphics[height=7.5cm, width=7.5cm]{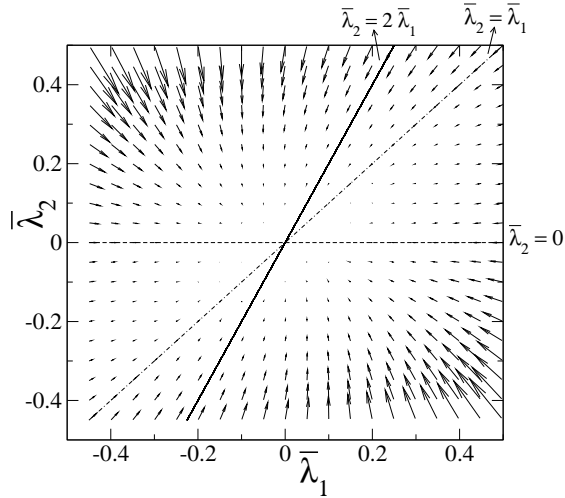}}
    \caption{RG flow diagram in the phase plane of dimensionless nonlinear couplings $\bar \lambda_{1}$ and $\bar \lambda_{2}$ defined in (\ref{lambdabars}). The solid line represents line $\bar \lambda_{2} = 2\bar \lambda_{1}$ (equilibrium limit), the dot dashed line represents $\bar \lambda_{1} = \bar \lambda_{2}$ and dashed line represents $\bar \lambda_{2} = 0$ (limit when equation is similar to Burgers equation). For these three cases, it is particularly easy to show analytically that the flow is inward (i.e. nonlinearities are marginally {\em irrelevant}). In fact for all $\bar \lambda_{1}$, $\bar \lambda_{2}$ the flow is towards (0, 0).}
    \label{flowdiag}
  \end{center}
\end{figure}

Here we calculate the recursion relation for all five parameters. From  (\ref{intermvariables}), the  constraint of rotational invariance $2(A_{1}-A_{2}) = \lambda_{2}$ requires
\begin{equation}
\xi(l) = \exp(\alpha(l) + 2l)\bigg(1+\frac{l G_{2}(\bar \lambda_{1}, \bar \lambda_{2})}{4\times 8\pi} + \frac{l F_{2}((\bar \lambda_{1}, \bar \lambda_{2})}{4\times 8\pi} \bigg)
\label{xidef}
\end{equation}
where the functions $G_{2}(\bar \lambda_{1}, \bar \lambda_{2})$ and $F_{2}(\bar \lambda_{1}, \bar \lambda_{2})$ are already defined in  (\ref{selfenergysymbols}) and (\ref{Flambda_interm}). With this choice of $\xi(l)$, substituting in  (\ref{intermvariables}), recursion relations for all five variables given by,
\begin{align}
 & \frac{dA_{1}}{dl} = A_{1}[-2+z(l) + \frac{G_{2}(\bar \lambda_{1}, \bar \lambda_{2})}{4\times 8\pi}],  \notag \\
 & \frac{dA_{2}}{dl} = A_{2}[-2+z(l) + \frac{G_{2}(\bar \lambda_{1}, \bar \lambda_{2})}{4\times 8\pi}], \notag \\
 & \frac{d\lambda_{1}}{dl} = \lambda_{1}[-2+z(l) + \frac{F_{2}(\bar \lambda_{1}, \bar \lambda_{2})}{4\times 8\pi} + \frac{G_{2}(\bar \lambda_{1}, \bar \lambda_{2})}{4\times 8\pi} \notag \\
 & - \frac{F_{1}(\bar \lambda_{1}, \bar \lambda_{2})}{4\times 8\pi}], \notag \\
 & \frac{d\lambda_{2}}{dl} = \lambda_{2}[-2+z(l) + \frac{G_{2}(\bar \lambda_{1}, \bar \lambda_{2})}{4\times 8\pi}], \notag \\
 & \frac{dD}{dl} = D[-2+z(l) + \frac{(\bar \lambda_{2} - \bar \lambda_{1})^{2}}{2 \times 8\pi} - \frac{G_{2}(\bar \lambda_{1}, \bar \lambda_{2})}{2\times 8\pi} \notag \\
 & - \frac{F_{2}(\bar \lambda_{1}, \bar \lambda_{2})}{2\times 8\pi}]. 
\label{recursionlation}
\end{align}
where $z(l)$ is defined by $\alpha(l) = \int_{0}^{l}z(l^{'})d l^{'}$, 
and the dimensionless variables $\bar \lambda_{1}$ and $\bar \lambda_{2}$ 
were defined in  (\ref{lambdabars}). 
The functions $G_{2}$, $F_{1}$ and $F_{2}$ 
are already defined in  (\ref{selfenergysymbols}) and (\ref{Flambda_interm}). 
In these recursion relations the function $z(l)$ is unknown at this point. 
It will drop out in the recursion relation for the dimensionless variables, 
$\bar \lambda_{1}$ and $\bar \lambda_{2}$, for which the recursion relations are
\begin{equation}
\frac{d\bar \lambda_{1}}{dl} = \bar \lambda_{1}\bigg[\frac{(\bar \lambda_{2} - \bar \lambda_{1})^{2}}{4 \times 8\pi} - \frac{3}{2}\frac{G_{2}(\bar \lambda_{1}, \bar \lambda_{2})}{4\times 8\pi} \bigg] -\frac{F_{1}^{*}(\bar \lambda_{1}, \bar \lambda_{2})}{4\times 8\pi}, 
\label{lambda1recursion}
\end{equation} 
\begin{equation}
\frac{d\bar \lambda_{2}}{dl} = \bar \lambda_{2}\bigg[\frac{(\bar \lambda_{2} - \bar \lambda_{1})^{2}}{4 \times 8\pi} - \frac{3}{2}\frac{G_{2}(\bar \lambda_{1}, \bar \lambda_{2})}{4\times 8\pi} -\frac{F_{2}(\bar \lambda_{1}, \bar \lambda_{2})}{4\times 8\pi}\bigg]. 
\label{lambda2recursion}
\end{equation}
Equations (\ref{lambda1recursion}) and  (\ref{lambda2recursion}) are coupled nonlinear equations for  $\bar \lambda_{1}$ and $\bar \lambda_{2}$. \\
In the special, high-symmetry case 
 $\lambda_{2} =0$, 
from  (\ref{selfenergysymbols}), 
$ G_{2}(\bar \lambda_{1}, \bar \lambda_{2}) = 2\lambda_{1}^{2}$ 
and $F_{1}(\bar \lambda_{1}, \bar \lambda_{2}) = 0$.
Then the  dimensionless coupling $\bar \lambda_{1}^{2}(l) = \lambda_{1}^{2}(l) D(l
)/A^{3/2}(l)$ obeys
\begin{equation}
\frac{d\bar \lambda_{1}}{dl} = \bar \lambda_{1}\bigg[-2+2- \frac{\bar \lambda_{1}^{2}(l)}{2\times 8\pi}\bigg] = -\frac{\bar
\lambda_{1}^{3}(l)}{2\times 8\pi}
\label{burgerlimitlambda}
\end{equation}
which tells us  $\bar \lambda_{1}$ is marginally {\em irrelevant}.
By contrast, for the Burgers equation in 2-d, the nonlinearity is marginally relevant. This is surprising, given the similarities of the two models in the limit $\lambda_2=0$.
A second special case is $\bar \lambda_{2} = 2\bar \lambda_{1}$, when the problem reduces to an equilibrium problem, as remarked in section \ref{eqmlimit}.
At this particular choice of 
$\bar \lambda_{1}$ and $\bar \lambda_{2}$, $G_{2}(\bar \lambda_{1}, \bar \lambda_{2}) = 0$, $(\bar \lambda_{2} - \bar \lambda_{1})^{2} = \bar \lambda_{1}^{2} = \frac{\bar \lambda_{2}^{2}}{4}$, $F_{1}(\bar \lambda_{1}, \bar \lambda_{2}) = 16\lambda_{1}^{3}$ 
and 
$F_{2}(\bar \lambda_{1}, \bar \lambda_{2}) = 4\bar \lambda_{2}^{2}$. 
Substituting these expressions for all functions in (\ref{lambda1recursion}) and  (\ref{lambda2recursion}), the flow equations for the equilibrium limit are
\begin{equation}
 \frac{d\bar \lambda_{1}}{dl} = -\frac{15}{4 \times 8\pi}\bar \lambda_{1}^{3};   \qquad
   \frac{d\bar \lambda_{2}}{dl} = -\frac{15}{4 \times 4 \times 8\pi}\bar \lambda_{2}^{3}   
\label{eqmlimitlambda}
\end{equation}   
We can draw the flow-diagram in $(\bar \lambda_{1}, \bar \lambda_{2})$ plane. (Fig {\ref{flowdiag}}) shows that for three special cases, $\lambda_{2}=0$, $\lambda_{2}=2\lambda_{1}$ and $\lambda_{2}=\lambda_{1}$ flow is towards zero. For other points also flow is towards zero. This means (0, 0) is the only fixed point and it is stable. We have checked this numerically as well. 

Since the nonlinearities are marginally irrelevant the effective stiffness $A_1$ and $A_2$ become equal at large scales, and are nonsingular. Therefore $<|\theta_{q}|^2> \sim q^{-2}$ for small ${\bf q}$, i.e. the renormalized theory still has only quasi long-ranged order.
\section{Conclusion and Discussion}
\label{discussion}
In this paper we have provided a systematic analysis of the large-scale, long-time behaviour of 
the stochastic nonlinear partial differential equation for the angle field of an active nematic on a 2-dimensional substrate. We constructed the general equation of motion for the order parameter, starting from a description that included the velocity, density as well. We then reduced the model to focus on the director or small-angle fluctuations about an ordered active nematic, and studied the evolution of the parameters therein under the dynamic renormalization group \cite{SMa, FNS, BurgerKPZ}. The  equation has five  parameters, 
$A_{1}$ and $A_{2}$ which are  
director diffusivities for two directions, the 
 nonlinear couplings  $\lambda_{1}$ and $\lambda_{2}$  and 
$D_{0}$ the noise strength. 
Two special cases are of interest: $\lambda_2=2\lambda_1$, for which the dynamics is that of an equilibrium two-dimensional nematic where static properties are shown to agree with \cite{NP}. The second case is $\lambda_2=0$, for which the equation can be mapped to a 
Burgers equation, 
for a  velocity field ${\bf v}$ given in (\ref{veffnem2d}), 
with
$\partial_{x}v_{x} - \partial_{z}v_{z} = 0$.
Despite this resemblance 
the dimensionless nonlinear coupling parameter 
$\bar \lambda^{2} = \frac{\lambda^{2}D_{0}}{A^{3}}$ 
is found to be marginally {\em irrelevant}, whereas for the  
Burgers equation in $d=2$ (see \cite{FNS}) the
nonlinearity was marginally {\em relevant}. 
Interestingly in this limit the 
diffusion constant and noise strength renormalize the same way, implying the  system has a hidden detailed balance, which we exposed via a Fokker-Planck analysis. 
The complete one-loop recursion relation for the five parameters constrained only by rotational-invariance show that the nonlinearities are always marginally irrelevant.\\ 
In Appendix \ref{appD} we present the equation of motion 
for the angle field starting from a velocity field which satisfies incompressibility. This provides another, inequivalent, situation in which the density is fast and can therefore be suitably eliminated. The procedure leads to a slightly different equation from (\ref{Qeomeff}) or (\ref{Qeomft}) with nonlocality due to transverse projectors. We have not analysed the properties of the incompressible version. Our results, despite the neglect of the density, are consistent with the numerical findings of \cite{chateginellimontagne}, that active nematic order in $d=2$ is quasi long-range. A complete treatment of the coupled behaviour of angle and density correlators in steady state, beyond the linearized analysis of \cite{sradititoner}, as well as a study of the incompressible model, are left for future work. 
\begin{acknowledgements}
SM thanks the CSIR, India for financial support.
SR acknowledges support from CEFIPRA project 3504-2, and from the DST, India
through the Centre for Condensed Matter Theory and Math-Bio Centre grant
SR/S4/MS:419/07
\end{acknowledgements}

\appendix
\section{Propagator renormalization}
\label{appA}
We start from the symmetrised 
version of (\ref{propagatorft}) (by substituting ${\bf k} \equiv \frac{{\bf q}}{2} + {\bf k}$ and $\Omega \equiv \frac{\omega}{2} + \Omega$)
\begin{align}
\Sigma ({\bf q}, \omega) & = 4 \times 2D_{0} \int_{k\Omega}M(\frac{{\bf q}}{2} + {\bf k}, \frac{{\bf q}}{2} - {\bf k}) \notag \\
                         & \times M(-\frac{{\bf q}}{2}-{\bf k}, {\bf q}) \times G_{0}(\frac{{\bf q}}{2} + {\bf k}, \frac{\omega}{2} + \Omega) \notag \\
                         & \times G_{0}(-\frac{{\bf q}}{2} - {\bf k}, -\frac{\omega}{2} -\Omega)G_{0}(\frac{{\bf q}}{2} - {\bf k}, \frac{\omega}{2} - \Omega)   
\label{symmetrizeselfenergy}
\end{align}  
where $ G_{0}({\bf q}, \omega) = (-i \omega + A_{1}q_{x}^{2} + A_{2}q_{z}^2)^{-1}$ is the unrenormalized propagator. It is easy to evaluate the $\Omega $ integral first in (\ref{symmetrizeselfenergy}). Separating the $\Omega$-integral
\begin{equation}
I_{\Omega}^{P}({\bf k}) = \int_{-\infty}^{+\infty} \bigg|G_{0}(\frac{{\bf q}}{2} + {\bf k}, \frac{\omega}{2} + \Omega)\bigg|^{2} G_{0}(\frac{{\bf q}}{2} - {\bf k}, \frac{\omega}{2} - \Omega) d\Omega 
\label{Omegaintegral}
\end{equation}
After substituting the expressions for the unrenormalized propagator in (\ref{Omegaintegral})
\begin{equation}
I_{\Omega}({\bf k}) = \int^{+\infty}_{-\infty} \frac{i(\frac{\omega}{2} - \Omega) + a}{[(\frac{\omega}{2} + \Omega)^{2} + b^{2}]\times [(\frac{\omega}{2} - \Omega)^{2} + a^{2}]} d\Omega
\label{Omegaint}
\end{equation}
where 
\begin{align}
& a =  [A_{1}(\frac{q_{x}}{2} - k_{x})^{2} + A_{1}(\frac{q_{z}}{2} - k_{z})^{2}] \notag \\
& b =  [A_{1}(\frac{q_{x}}{2} + k_{x})^{2} + A_{1}(\frac{q_{z}}{2} + k_{z})^{2}]  
\label{defab}
\end{align}
After integrating $I_{\Omega}({\bf k})$ over $\Omega$, for $\omega \longrightarrow 0$, we see that,
\begin{equation}
 I_{\Omega}({\bf k}) = \frac{\pi}{b(a+b)}.
\label{Iomega}
\end{equation} 
Substituting this $\Omega$ integral in the calculation of the self-energy (\ref{symmetrizeselfenergy}) 
\begin{align}
\Sigma ({\bf q}, \omega) & = 4 \times 2D_{0} \pi \frac{1}{(2\pi)^{2+1}}\int M(\frac{{\bf q}}{2} + {\bf k}, \frac{{\bf q}}{2} - {\bf k}) \notag \\
                         &   M(-\frac{{\bf q}}{2}-{\bf k}, {\bf q}) \times
                          \frac{1}{b(a+b)} d{\bf k},    
\label{selfenergykint}
\end{align}  
where $a$ and $b$ are defined in  (\ref{defab}). Since we are interested in long-wavelength properties, we can do small $q_{x}$ and $q_{y}$ expansions. For calculating $\Sigma ({\bf q}, \omega)$, we need to perform the ${\bf k}$ integral. Defining small parameters $x=\frac{q_{x}}{k_{x}}$ and $z=\frac{q_{z}}{k_{z}}$, and expanding up to lowest order in $x$ and $z$
\begin{align}
\frac{1}{b(a+b)} & = \frac{1}{2 k_{x}^{4} \alpha^{2}}\bigg[ 1 - \frac{x^{2}}{2\alpha}A_{1} - \frac{z^{2}}{2\alpha}A_{2}\tan^{2}\theta \notag\\
                 &  - \frac{x}{\alpha}A_{1} - \frac{z}{\alpha}A_{2}\tan^{2}\theta + \frac{x^{2}}{\alpha^{2}} A_{1}^{2}\notag \\
                 &  + \frac{z^{2}}{\alpha^{2}} A_{2}^{2}\tan^{4}\theta + \frac{2xz}{\alpha^{2}} A_{1}A_{2}\tan^{2}\theta\bigg]       
\label{defbaplusb}
\end{align}
where $\theta = \tan^{-1}(\frac{k_{z}}{k_{x}})$ and $\alpha = (A_{1} + A_{2} \tan^2\theta) $. The next step for the calculation of the integral is the product of two propagators $M \times M$ in (\ref{selfenergykint}). 
\begin{align}
 & M(\frac{{\bf q}}{2} + {\bf k}, \frac{{\bf q}}{2} - {\bf k}) \times M(-\frac{{\bf q}}{2}-{\bf k}, {\bf q}) \notag  \\
                                                                                                  & = \frac{k_{x}^{2}k_{z}^{2}}{4}\bigg[xz G_{1}(\lambda_{1}, \lambda_{2}) + (x+z)G_{2}(\lambda_{1}, \lambda_{2}) \notag \\ 
                                                                                                  & + 2 G_{3}(\lambda_{1}, \lambda_{2})\bigg] 
\label{productMM}
\end{align}
From (\ref{defbaplusb}) and  (\ref{productMM}) integrand of (\ref{selfenergykint}) is,
\begin{align}
& \frac{ M(\frac{{\bf q}}{2} + {\bf k}, \frac{{\bf q}}{2} - {\bf k}) \times M(-\frac{{\bf q}}{2}-{\bf k}, {\bf q})}{b(a+b)} \notag \\
& = \frac{k_{x}^{2}k_{z}^{2}}{4 \times 2 k_{x}^{4} \alpha^{2}} \bigg[ x z G_{1} \notag + G_{2}\bigg(- \frac{x^{2}}{\alpha}A_{1} \notag\\
&  - \frac{z^{2}}{\alpha}A_{2}\tan^{2}\theta - \frac{x z}{\alpha}A_{1} - \frac{x z}{\alpha}A_{2}\tan^{2}\theta\bigg) + 2 G_{3}\notag \\
&  \bigg( 1 - \frac{x^{2}}{2\alpha}A_{1} - \frac{z^{2}}{2\alpha}A_{2}\tan^{2}\theta  - \frac{x}{\alpha}A_{1} - \frac{z}{\alpha}A_{2}\tan^{2}\theta \notag \\
& + \frac{x^{2}}{\alpha^{2}} A_{1}^{2} +  \frac{z^{2}}{\alpha^{2}} A_{2}^{2}\tan^{4}\theta + \frac{2xz}{\alpha^{2}} A_{1}A_{2}\tan^{2}\theta \bigg) \bigg]       
\label{totalkintegrad}
\end{align}
On integration (inside the $[\qquad]$) only term of $O(x^{2})$, of  $O(z^{2})$ and  $O(1)$ survive. Hence terms which will contribute to the integration are
\begin{align}
& G_{2}\bigg(-\frac{x^{2}}{\alpha}A_{1} - \frac{z^{2}}{\alpha}A_{2}\tan^{2}\theta \bigg) \notag \\
& + 2 G_{3} \bigg( 1 - \frac{x^{2}}{2\alpha}A_{1} - \frac{z^{2}}{2\alpha}A_{2}\tan^{2}\theta + \frac{x^{2}}{\alpha^{2}} A_{1}^{2} \notag \\
& +  \frac{z^{2}}{\alpha^{2}} A_{2}^{2}\tan^{4}\theta \bigg) 
\label{finalkintegrand}
\end{align}
where  $ G_{2} = (2\lambda_{2}^{2} + \lambda_{2}^{2} - 3\lambda_{1}\lambda_{2})$ and $G_{3} = (\lambda_{2}^{2} - \lambda_{1}\lambda_{2})$. $k_{x} = k\cos \theta $ and $k_{z} = k\sin \theta $ and $\alpha = (A_{1} + A_{2}\tan^{2}\theta)$. After performing the integration for these two types of terms  in (\ref{finalkintegrand}),
\begin{align}
\Sigma ({\bf q}, \omega \rightarrow 0) & = \frac{l}{4\pi}\bigg [-\frac{G_{2}(\bar \lambda_{1}, \bar \lambda_{2})}{8}(A_{1}q_{x}^{2} + A_{2}q_{y}^{2}) \notag \\
                         & + \frac{G_{3}(\bar \lambda_{1}, \bar \lambda_{2}) A_{1}A_{2}}{(\sqrt{A_{1}} + \sqrt{A_{1}})^{2}} \bigg] 
\label{selfenergyomegazero}
\end{align}    
This is the expression for the self-energy as given in (\ref{eqselfenergy}).
\section{Vertex renormalization}
\label{appB}
Here we calculate the three-point symmetrised vertex function $\Gamma$.
There are three distinct one-loop diagrams 
$\Gamma_{a}$, $\Gamma_{b}$ and $\Gamma_{c}$ 
contributing to the correction to the vertex as shown in 
(Fig \ref{perttheoryfig}(b)). These diagrams all have multiplicity 4. 
In this Appendix we will go into the details of the calculation of $\Gamma_{a}$. 
The calculations for $\Gamma_{b}$ and $\Gamma_{c}$ are the same as for 
$\Gamma_{a}$. Small variables $x$ and $z$ are  as defined in 
Appendix \ref{appA}: for self-energy. 
We start from the symmetrised version of (\ref{eqGamma_a})
\begin{align}
\Gamma_{a}({\bf q}, {\bf k_{1}}) & = 4 \times 2D_{0} \int_{k\Omega}M(\frac{{\bf q}}{2} + {\bf k}, \frac{{\bf q}}{2} - {\bf k}) \notag \\
                                 &   \times M(\frac{{\bf q}}{2} + {\bf k}_{1}, {\bf k} - {\bf k}_{1}) \notag \\
                                 &   \times M(\frac{{\bf q}}{2} -  {\bf k_{1}}, -{\bf k} + {\bf k_{1}})  \notag \\
                                 &   \times \bigg|G_{0}({\bf k} - {\bf k}_{1}, \Omega - \Omega_{1})\bigg|^{2} \times \notag \\
                                 &   G_{0}(\frac{{\bf q}}{2} + {\bf k}, \frac{\omega}{2} + \Omega) \times G_{0}(\frac{{\bf q}}{2} - {\bf k}, \frac{\omega}{2}-\Omega) 
\label{Gammaa_symmetrized}
\end{align}
Separating the $\Omega$ integral part from the full integration in (\ref{Gammaa_symmetrized})
\begin{align}
I^{V}_{a \Omega}({\bf k}) & = \int^{+\infty}_{-\infty} \bigg|G_{0}({\bf k} - {\bf k}_{1}, \Omega - \Omega_{1})\bigg|^{2} \times \notag \\
                                 &   G_{0}(\frac{{\bf q}}{2} + {\bf k}, \frac{\omega}{2} + \Omega) G_{0}(\frac{{\bf q}}{2} - {\bf k}, \frac{\omega}{2}-\Omega)d\Omega  
\label{IVa}
\end{align}
for $\omega \longrightarrow 0$ and $\Omega_{1} \longrightarrow 0$ limit and writing in terms of real and imaginary parts,
\begin{equation}
Re(I^{V}_{a \Omega}({\bf k}))  = \int \frac{ab + \Omega^{2}}{(\Omega^{2}+b^{2})(\Omega^{2}+a^{2})(\Omega^{2}+c^{2})}
\label{IVa1}
\end{equation}
 For $\omega \longrightarrow 0$ and $\Omega_{1} \longrightarrow 0$ 
limits $Im(I^{p}_{\Omega}({\bf k})) = 0$. 
where $a$ and $b$ are as defined in (\ref{defab}), and 
\begin{equation}
c =  [A_{1}(k_{x} - k_{x_{1}})^{2} + A_{1}(k_{z} - k_{z_{1}})^{2}] 
\label{defc}
\end{equation}
Performing the integral over $\Omega$, 
\begin{equation}
 I^{V}_{a \Omega}({\bf k}) = \frac{\pi (2c + a + b)}{c(a + c)(b + c)(a + b)} 
\label{IVafinal}
\end{equation}
Similarly for $\Gamma_{b}$ and  $\Gamma_{c}$,
\begin{align}
& I^{V}_{b \Omega}({\bf k}) = \frac{\pi}{a(a + c)(a + b)} \notag \\
& I^{V}_{c \Omega}({\bf k}) = \frac{\pi}{b(b + c)(a + b)}  
\label{IVbc}
\end{align}
Substituting this $I^{V}_{a \Omega}({\bf k})$ from (\ref{IVafinal}) in the calculation of  $\Gamma_{a}$,
\begin{align}
\Gamma_{a}({\bf q}, {\bf k_{1}}) & = 4 \times 2D_{0} \int_{k\Omega}M(\frac{{\bf q}}{2} + {\bf k}, \frac{{\bf q}}{2} - {\bf k})  \notag \\
                                 &   \times M(\frac{{\bf q}}{2} + {\bf k}_{1}, {\bf k} - {\bf k}_{1}) \notag \\
                                 &   \times M(\frac{{\bf q}}{2} -  {\bf k_{1}}, -{\bf k} + {\bf k_{1}})  \notag \\
                                 &   \times \frac{\pi (2c + a + b)}{c(a + c)(b + c)(a + b)} 
\label{Gamma_a_int}
\end{align}
We are interested in long wavelength properties. By defining the small quantities $x=\frac{q_{x}}{k_{x}}$, $z=\frac{q_{z}}{k_{z}}$, $x_{1}=\frac{k_{x_{1}}}{k_{x}}$ and  $z_{1}=\frac{k_{z_{1}}}{k_{z}}$, where $k_{x} = k \cos \theta$ and  $k_{z} = k \sin \theta$, up to lowest order in $x$, $z$, $x_{1}$ and $z_{1}$,
\begin{align}
I^{V}_{a \Omega}({\bf k}) & = \frac{\pi (2c + a + b)}{c(a + c)(b + c)(a + b)} \notag \\
                          & = \frac{\pi}{2 k_{x}^{6} \alpha^{3}}\bigg[1 + \frac{3 x_{1}}{\alpha}A_{1} + \frac{3 z_{1}}{\alpha}A_{2}\tan^{2}\theta \notag \\
                          & + \frac{x z }{2\alpha^{2}}A_{1}A_{2}\tan^{2}\theta + \frac{14 x_{1} z_{1}}{\alpha^{2}}A_{1}A_{2}\tan^{2}\theta \bigg]  
\label{Gamma_a_final}
\end{align}   
The next step for the calculation of the integral is the product of three propagators $M \times M \times M$
\begin{align}
 & M(\frac{{\bf q}}{2} + {\bf k}, \frac{{\bf q}}{2} - {\bf k})\times M(\frac{{\bf q}}{2} + {\bf k}_{1}, {\bf k} - {\bf k}_{1}) \notag  \\ 
 & \times M(\frac{{\bf q}}{2} -  {\bf k_{1}}, -{\bf k} + {\bf k_{1}}) \notag \\
 & = 2k_{x}^{3} k_{z}^{3} \bigg[\bigg(\frac{\lambda_{1}}{2}\bigg)^{3}\bigg( 2(\frac{xz}{4} - x_{1} z_{1}) \bigg ) \notag\\
 & + \bigg(\frac{\lambda_{1}}{2}\bigg)^{2}\bigg(\frac{\lambda_{2}}{2}\bigg)\bigg( -\frac{xz}{2} + 10 x_{1} z_{1} - 2 x_{1} -2 z_{1}\bigg ) \notag \\ 
 & +  \bigg(\frac{\lambda_{2}}{2}\bigg)^{2}\bigg(\frac{\lambda_{1}}{2}\bigg)\bigg(-\frac{xz}{4}-14 x_{1} z_{1} + 4 x_{1} + 4 z_{1} -1\bigg) \notag \\
 & +  \bigg(\frac{\lambda_{2}}{2}\bigg)^{3}\bigg( \frac{3xz}{4} + 6 x_{1} z_{1} - 2 x_{1} -2 z_{1} +1\bigg) \bigg]     
\label{MMM}
\end{align}
From  (\ref{Gamma_a_final}) and  (\ref{MMM}), the  product inside the integral for $\Gamma_{a}({\bf q}, {\bf k_{1}})$ is 
\begin{align}
& \frac{\pi (2c + a + b)}{c(a + c)(b + c)(a + b)} \times M(\frac{{\bf q}}{2} + {\bf k}, \frac{{\bf q}}{2} - {\bf k}) \notag  \\
& M(\frac{{\bf q}}{2} + {\bf k}_{1}, {\bf k} - {\bf k}_{1})\times M(\frac{{\bf q}}{2} -  {\bf k_{1}}, -{\bf k} + {\bf k_{1}}) \notag \\
& = \frac{ \pi 2k_{x}^{3} k_{z}^{3}}{2 k_{x}^{6} \alpha^{3}} \bigg[ 2\bigg(\frac{\lambda_{1}}{2}\bigg)^{3}\bigg( 2(\frac{xz}{4} - x_{1} z_{1}) \bigg )\notag \\
& + \bigg(\frac{\lambda_{1}}{2}\bigg)^{2}\bigg(\frac{\lambda_{2}}{2}\bigg)\bigg(-\frac{xz}{2} + 10 x_{1} z_{1} - \frac{6 x_{1} z_{1}}{\alpha}A_{1} \notag \\
& - \frac{6 x_{1} z_{1}}{\alpha}A_{2} \tan^{2}\theta\bigg) \notag \\
& + \bigg(\frac{\lambda_{2}}{2}\bigg)^{2}\bigg(\frac{\lambda_{1}}{2}\bigg)\bigg(-\frac{xz}{4} - 14 x_{1} z_{1} + \frac{12 x_{1} z_{1}}{\alpha}A_{1} \notag \\
& + \frac{12 x_{1} z_{1}}{\alpha}A_{2} \tan^{2}\theta - \frac{x z}{2 \alpha^{2}} A_{1}A_{2} \tan^{2}\theta  \notag \\
& - \frac{14 x_{1} z_{1}}{\alpha^{2}} A_{1}A_{2} \tan^{2}\theta \bigg) \notag \\
& + \bigg(\frac{\lambda_{2}}{2}\bigg)^{3}\bigg( \frac{3xz}{4} + 6 x_{1} z_{1} - \frac{6 x_{1} z_{1}}{\alpha}A_{1} \notag \\
& - \frac{6 x1 z1}{\alpha}A_{2} \tan^{2}\theta + \frac{x z}{2 \alpha^{2}} A_{1}A_{2} \tan^{2}\theta  \notag \\
& + \frac{14 x_{1} z_{1}}{\alpha^{2}} A_{1}A_{2} \tan^{2}\theta \bigg)    
\label{total_int_Gamma}
\end{align}
We display only those terms which give a nonzero contribution after integrating over ${\bf k}$. Similarly we can obtain expressions for $\Gamma_{b}$ and $\Gamma_{c}$ 

The total $\Gamma = \Gamma_{a} + \Gamma_{b} + \Gamma_{c} =  \Gamma_{a} + 2 \Gamma_{b}$. After doing the integration over ${\bf k}$, the final expression for $\Gamma$,
\begin{align}
\Gamma({\bf q}, {\bf k_{1}}) & = 2 (k_{x}k_{z})\frac{1}{2}\bigg[\bigg(\frac{\lambda_{1}}{2}\bigg)^{2}\bigg(\frac{\lambda_{2}}{2}\bigg)\bigg( -\frac{x z}{16\pi}  \notag \\
                             & + \frac{x_{1} z_{1}}{4\pi}\bigg) + \bigg(\frac{\lambda_{2}}{2}\bigg)^{2}\bigg(\frac{\lambda_{1}}{2}\bigg)\bigg( -\frac{x z}{32\pi}  \notag \\
                             &  - \frac{7 x_{1} z_{1}}{8 \pi} \bigg) + \bigg(\frac{\lambda_{2}}{2}\bigg)^{3}\bigg( -\frac{7x z}{32 \pi} + \frac{x_{1} z_{1}}{8\pi}\bigg) \bigg] 
\label{full_Gamma}
\end{align}    
The bare vertex is
\begin{align}
\Gamma_{0}({\bf q}, {\bf k_{1}}) & = 2 (k_{x}k_{z}) \bigg[ \frac{\lambda_{1}}{2}\bigg(\frac{x z}{4} - x_{1} z_{1}) \bigg) \notag  \\
                                 & + \frac{\lambda_{2}}{2}\bigg(\frac{x z}{4} + x_{1} z_{1}) \bigg)\bigg]  
\label{barevertex}
\end{align}
Decomposing expression in (\ref{barevertex}) into parts of the form $(\frac{x z}{4} - x_{1} z_{1}) $ and $(\frac{x z}{4} + x_{1} z_{1}) $, we get the corrections to $
\frac{\lambda_{1}}{2}$ and $\frac{\lambda_{2}}{2}$. Hence with this  decomposition (\ref{barevertex}) can be rewritten as
\begin{align}
\Gamma({\bf q}, {\bf k_{1}}) & = 2 (k_{x}k_{z})\bigg(\frac{x z}{4} - x_{1} z_{1}\bigg) \bigg[-\frac{\lambda_{1}^{2}\lambda_{2}}{4 \times 2 \times8 \pi} \notag \\
                             & + \frac{3 \lambda_{2}^{2}\lambda_{1}}{2 \times 8\times8 \pi} +  \frac{\lambda_{2}^{3}}{2 \times 8\times8 \pi} \bigg] \notag \\
                             & + 2 (k_{x}k_{z})\bigg(\frac{x z}{4} + x_{1} z_{1}\bigg)  \notag \\
                             & \bigg[ -\frac{4 \lambda_{2}^{2}\lambda_{1}}{2 \times 8\times8 \pi} + \frac{6\lambda_{2}^{3}}{2 \times 8\times8 \pi} \bigg ]  
\label{Gamma}
\end{align}
Comparing with the expression for the original vertex, the 
corrections to $\frac{\lambda_{1}}{2}$ and $\frac{\lambda_{2}}{2}$ are
\begin{align}
 & \tilde{\lambda}_{1} = \lambda_{1}\bigg[1 - \frac{F_{1}(\bar \lambda_{1}, \bar \lambda_{2})l}{2\times 8\pi}\bigg] \notag \\
 & \tilde{\lambda}_{2} = \lambda_{2}\bigg[1 - \frac{F_{2}(\bar \lambda_{1}, \bar \lambda_{2})l}{2\times 8\pi}\bigg]  
\label{lambdatilde}
\end{align}
where functions $F_{1}(\bar \lambda_{1}, \bar \lambda_{2})$ and $F_{2}(\bar \lambda_{1}, \bar \lambda_{2})$ are defined in (\ref{Flambda_interm})
\section{Noise strength renormalization}
\label{appC}
Here we will compute the leading-order correction to the noise strength. The relevant diagram which will contribute to the integral is shown in (Fig \ref{perttheoryfig}(c)); it has multiplicity of 2. Calculating the integral with this symmetrised vertex,
\begin{align}
\Delta {D} & = 2 \times (2D_{0})^{2} \int_{k\Omega} M(\frac{{\bf q}}{2} + {\bf k}, \frac{{\bf q}}{2}- {\bf k}) \notag \\
           &   M(-\frac{{\bf q}}{2} -{\bf k}, {\bf k}- \frac{{\bf q}}{2}) \bigg|G_{0}(\frac{{\bf q}}{2}+{\bf k}, \frac{\omega}{2}+\Omega)\bigg|^{2} \bigg| \notag \\
           & G_{0}(\frac{\bf q}{2} - {\bf k}, \frac{\omega}{2}-\Omega)\bigg|^{2} 
\label{DeltaD}
\end{align} 
Separating the $\Omega$ integral from the full integration and taking $\omega \longrightarrow 0$,
\begin{equation}
I^{D}_{\Omega}{\bf k} = \frac{\pi}{ab(a+b)}
\label{ID}
\end{equation}  
Expanding  $\frac{1}{ab(a+b)}$ as in the calculation of the propagator in terms of small variables $x$ and $z$, the  terms which will contribute to lowest order are of order 1. Hence to lowest order,
\begin{equation}
\frac{1}{ab(a+b)} \simeq \frac{1}{2 k_{x}^{6} \alpha^{3}}
\label{Dab}
\end{equation}
The next step of the  calculation of the integral is the product of two propagators, $M\times M$. To lowest order,
\begin{equation}
M(\frac{{\bf q}}{2} + {\bf k}, \frac{{\bf q}}{2}- {\bf k})M(-\frac{{\bf q}}{2} -{\bf k}, {\bf k}- \frac{{\bf q}}{2}) = k_{x}^{2}k_{z}^{2} (\lambda_{2}- \lambda_{1})^{2}    
\label{MMD}
\end{equation}
The final expression for the product
\begin{align}
& \frac{1}{ab(a+b)}\times M(\frac{{\bf q}}{2} + {\bf k}, \frac{{\bf q}}{2}- {\bf k})M(-\frac{{\bf q}}{2} -{\bf k}, {\bf k}- \frac{{\bf q}}{2}) \notag \\
&  = \frac{k_{x}^{2}k_{z}^{2} (\lambda_{2}- \lambda_{1})^{2}}{2 k_{x}^{6} \alpha^{3}}
\label{integrandD}
\end{align}
After performing the integration over ${\bf k}$ in the integral (\ref{DeltaD}),
\begin{equation}
\Delta {D} = \frac{D_{0}^{2} (\lambda_{2}- \lambda_{1})^{2} l}{8 \pi (A_{1} A_{2})^{3/2}}
\label{DeltaDfinal}
\end{equation}
This gives
\begin{equation}
\tilde{D} =  D_{0}\bigg[ 1 + \frac{(\bar \lambda_{2} - \bar \lambda_{1})^{2} l}{2\times 8\pi}\bigg]
\label{tildeD}
\end{equation}    
\section{An incompressible active nematic}
\label{appD}
In this section we give the equation for the angle field  
 $\theta$, 
obtained from an incompressible  velocity field ${\bf v}$
($\nabla \cdot {\bf v} = 0$). From  (\ref{eqmomgen}), imposing $\rho = \mbox{constt}$ and $\nabla \cdot {\bf v} =0$, and defining the transverse projector $P = ({\bf 1} - \hat {\bf q} \hat {\bf q})$, we see that 
\begin{equation}
{\bf v} = -\bar \Gamma^{-1}   P \cdot (\nabla \cdot {\Q})
\label{incompressiblevdef}
\end{equation}
writing ${\Q}$ in terms of $\theta$
\begin{equation}
{\bf v} = -\bar \Gamma^{-1}  P \cdot (\partial_{z}\theta, \partial_{x}\theta)
\label{incompressiblevtheta}
\end{equation}
Substituting the expression for ${\bf v}$ in  (\ref{Qeom}) to linear order in $\theta$ the equation of motion 
\begin{align}
\bar G_{0}^{-1}({\bf q}, \omega) \theta_{{\bf q}, \omega} & = f_{\theta}({\bf q}, \omega) - \int_{{\bf k}, \Omega} \theta_{{\bf k}, \Omega} \theta_{{{\bf q} - {\bf k}}, \omega-\Omega} \bigg[ \bigg( \gamma_{1} \notag  \\
                                                         & - \frac{\alpha_{0}}{2}[ P_{22}(\hat {\bf k})+  P_{22}(\hat {\bf q} - \hat {\bf k}) -  P_{11}(\hat {\bf k}) \notag \\ 
                                                    &  -  P_{11}(\hat {\bf q} - \hat {\bf k})] \bigg) \times \bigg( M({\bf k}, {\bf q}-{\bf k}) \bigg) \notag \\
                                                    & +  \gamma_{2}  \bigg([  P_{12}(\hat {\bf k}) +  P_{12}(\hat {\bf q} - \hat {\bf k})] {\bf k}\cdot ({\bf q} - {\bf k})\notag \\
                                                    & + [ P_{11}(\hat {\bf k}) +  P_{22}(\hat {\bf q} - \hat {\bf k})] k_{y} (q_{x}-k_{x}) \notag \\
                                                    & + [ P_{22}(\hat {\bf k}) +  P_{11}(\hat {\bf q} - \hat {\bf k})] k_{x} (q_{y}-k_{y}) \bigg)\bigg]                 
\label{finaleqincompressible}
\end{align}
where $f_{\theta}({\bf q}, \omega)$ is Gaussian random nonconserving noise with 
noise-noise correlation as defined in (\ref{Qnoisecorrft}). $\bar G_{0}^{-1}({\bf q}, \omega)$ is inverse propagator, defined by
\begin{align}
\bar G_{0}^{-1}({\bf q}, \omega) & = \bigg(-i \omega + \frac{\alpha_{0}}{2} {\bf q}^{2} \notag \\
                                 & + A_{1} P_{11}(\hat {\bf q})q_{z}^{2} - A_{2} P_{22}(\hat {\bf q})q_{x}^{2} \bigg)^{-1}  
\label{barepropagatorincompressible}
\end{align}
$M({\bf k}, {\bf q}-{\bf k})$ as defined in  (\ref{Qvtxft}), 
 $ P_{11}(\hat {\bf q})$, $ P_{22}(\hat {\bf q})$ are diagonal components and $ P_{12}(\hat {\bf q})$ is the off-diagonal component of projection operator. We have not studied further the properties of this equation.


{}


\begin{thebibliography}{}
\bibitem{temperaturefootnote} For simplicity we consider a strictly isothermal system, so that the energy density or temperature field can be ignored.
\bibitem{multimplicativenoisefootnote}For simplicity we ignore the dependence of the noise strength on the dynamical variables. This would give rise to multiplicative noise effects that are beyond the scope of this work.
\bibitem{collisionalderivation} If we derive the equation for $\theta$ from a collisional model, where each particle moves forward or backward along its length and where two particles which come within a certain radius of each other try to align parallel to each other, we find
\begin{center}
$\frac{\partial \theta}{\partial t}=
\lambda_1(\theta)\partial_{x}\theta \partial_{z}\theta+
A_{1}(\theta)\partial_{x}^{2}\theta+
A_{2}(\theta)\partial_{z}^{2}\theta+
\lambda_2 \theta\partial_{x}\partial_{z}\theta$
\end{center}
upto quadratic order in $\theta$ and gradients, and ignoring  derivatives of the density. The coefficients are
\begin{center}
$ \lambda_1(\theta)=\frac{8}{3}S^{2}\cos2\theta-\frac{1}{3}S(1-\cos4\theta) \simeq \frac{8}{3}S^2, $
\end{center}
\begin{center} 
$ A_{1}(\theta)=-\frac{1}{2}S^{2}\cos2\theta+\frac{1}{6}S-\frac{1}{3}S\cos4\theta \simeq -\frac{1}{2}S^2-\frac{1}{6}S,$ 
\end{center}
\begin{center}
$ A_{2}(\theta)=\frac{1}{2}S^{2}\cos2\theta+\frac{1}{6}S-\frac{1}{3}S\cos4\theta \simeq \frac{1}{2}S^2-\frac{1}{6}S,$
\end{center}
\begin{center} 
$ \lambda_2 \simeq 2S^2.$
\end{center}
where $S$ is the scalar order parameter. Comparing with  (\ref{rotinv}), we see it satisfy the relation $2(A_2-A_1)=\lambda_2$.
\bibitem{noneqmtermfootnote}This can be confirmed by checking that the functional curl is nonzero:
\bibitem{comparisonburgerfootnose} This term is absent in the Burgers-like  limit $\lambda_2 =0$.
\end{thebibliography}
\end{document}